\documentclass[aps,pra,showpacs,twocolumn,groupedaddress]{revtex4}
\usepackage{graphicx}
\usepackage{epstopdf}
\usepackage{amsmath}
\usepackage{subfigure}
\usepackage{siunitx}
\usepackage{color}

\definecolor{blue1}{rgb}{0.05,0.1,0.6}

\usepackage[%
				   breaklinks=true,%
                   colorlinks=true,%
                   linkcolor=blue1,%
                   citecolor=blue1,%
                   urlcolor=blue1,%
                 ]{hyperref} 

\usepackage{amsfonts}

\begin{document}

\title{Observation of Subradiant Atomic Momentum States with Bose-Einstein Condensates in a Recoil Resolving Optical Ring Resonator}
\author{P. Wolf, S. C. Schuster, D. Schmidt, S. Slama, C. Zimmermann}
\affiliation{Physikalisches Institut, Eberhard-Karls-Universit\"{a}t T\"{u}bingen, Auf
der Morgenstelle 14, D-72076 T\"{u}bingen, Germany.}

\begin{abstract}
We experimentally investigate the formation of subradiant atomic momentum states in Bose-Einstein condensates inside a recoil resolving optical ring resonator according to the theoretical proposal of Cola, Bigerni, and Piovella. The atoms are
pumped from the side with laser light that contains two frequency components.
They resonantly drive cavity assisted Raman transitions between three
discreet atomic momentum states. Within a few hundred microseconds, the system
evolves into a stationary subradiant state. In this state, the condensate develops two density gratings suitable to diffract the two frequency components of the pump field into the resonator. Both components destructively interfere such that scattering is efficiently suppressed. A series of subradiant states for various amplitude 
ratios of the two pump components between 0 and 2.1 have been observed. The results are well explained with a three state quantum model in mean field approximation.  
\end{abstract}

\maketitle

Predicted by Dicke already in 1954 \cite{Dicke}, an optical excitation
inside an atomic cloud can decay faster or slower as compared to a single
excited atom \cite{Haroche}. While Dicke-superradiance has been experimentally
investigated in many systems, experimental examples for Dicke-subradiance are
rare. This is because superradiant states can be directly excited with an
optical transition, while subradiant states of two level systems do not
couple to the environment by radiation and need to be engineered with
appropriate methods. However, it is just this decoupling from the
environment and the resulting robustness against decoherence \cite{Guhne}
that makes subradiant systems interesting for storage of quantum
information \cite{Briegel} and for quantum information processing \cite%
{Sangouard}. A first experimental hint for subradiance was found already in
1985 \cite{CrubellierPRL}. Since then, subradiant type suppression
was observed in specially designed scenarios with two ions \cite{DeVoe}%
, two Cooper pair boxes \cite{Walraff}, and two modes in a single plasmonic
nanocavity \cite{Maier}. First, subradiance in a large ensemble of emitters
has been experimentally demonstrated only recently with cold thermal atoms
of a magneto-optical trap \cite{Kaiser}, with atoms coupled to a nonofiber \cite{Solano} and in arrays of plasmonic antennas \cite{Jenkins}. Instead of coupling the radiation to the states of the electronic shell, it
 is also possible to use the momentum
states of the atomic center of mass motion and couple them via spontaneous Raman transitions. As shown in pioneering experiments with
Bose-Einstein condensates \cite{Ketterle}, this Raman scattering is superradiant in a
cloud of atoms that collectively interact with the same light field. The superradiant behavior can be controlled by optical resonators tuning them in and out of resonance with respect to the frequency of the scattered
photons \cite{Bux},\cite{Hemmerich}. The analog subradiant suppression of spontaneous Raman scattering has not yet been observed but there is a
proposal \cite{Piovella} that discusses subradiance in systems
of three atomic momentum states. Different from two level systems, three level systems
are expected to spontaneously decay into subradiant states \cite%
{Crubellier}, which elegantly solves the problem of preparation (see Supplemental Material \cite{supplement_mat}).
Furthermore, the subradiant states can be manipulated by controlling the relative transition rates between the three states. In this Letter we describe the experimental realization of this proposal and present the observation of subradiant states in three level ladder systems that are made of momentum excitations in Bose-Einstein condensates. 

To explain the used scheme we first regard a single atom at rest inside an optical ring resonator. The atom is illuminated from the side by a laser pump beam. By scattering a photon from the pump beam into the cavity the atom is recoil kicked and undergoes a transition into a state with momentum $\vec{q}$. The frequency of the  scattered photon is slightly redshifted relative to the pump beam due to the kinetic energy that is transferred to the atom. The high finesse resonator resolves this shift and supports the scattering only if tuned to the frequency of the scattered photon. During the decay, the atom is in a superposition of two momentum states and thus forms a periodic spatial probability distribution. This fulfills the Bragg condition for optical diffraction of light from the pump beam into the propagating optical mode of the resonator. The photon in the optical mode eventually escapes the resonator through one of the mirrors. Apparently, the atom forms a two level system that may decay from an initial zero momentum state into a final state with finite momentum. With the atom being replaced by an atomic Bose-Einstein condensate the light scattered by the atoms coherently adds up such that the scattered intensity scales with $N^{2}$. The collective decay is thus superradiant. 

One may now add a third atomic state with momentum $2\vec{q}$. It is coupled to the state with momentum $\vec{q}$ by means of a second frequency component in the pump beam.  One would now expect the atoms to completely decay in a cascade from the initial zero momentum state to the final state with momentum $2\vec{q}$. The analysis in \cite{Piovella}, however, shows that the system decays into a stationary superposition of all three momentum states. This highly entangled collective atomic state still contains excitation and exhibits a time dependent periodic spatial density modulation. However, the two diffracted frequency components of the pump beam now interfere destructively such that no light is scattered from the pump beam into the resonator. The suppression of scattering by collective destructive interference is typical for subradiant systems.  

   \begin{figure}[htb]
        \begin{center}
            \includegraphics[width=3.3607in]{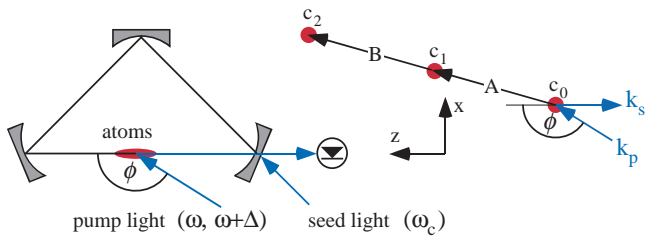}
        \end{center}
 \caption {Scheme for matter wave subradiance.  The three level system of atomic momentum states 
is indicated as red dots in momentum space (right subplot). Pump light with two components of frequency $\omega $ and $\omega
 +\Delta $ is scattered into the same optical mode of a recoil resolving ring
 resonator.}           
    \end{figure}

Figure 1 shows the experimental setup in more detail and introduces the relevant notions.
A pump beam with frequency $\omega $ and wave vector $\vec{k}_{p}$
illuminates the condensate from the side with an angle $\phi $ relative to
the optical axis of the cavity ($z$ direction in FIG. 1). The atoms scatter
photons into the two degenerate cavity modes that circulate in opposite
directions. Each scattering event changes the momentum of the atom by either 
$\hbar \vec{q}_{1}=\hbar \left( \vec{k}_{p}+\vec{k}_{s}\right) $ or $\hbar 
\vec{q}_{2}=\hbar \left( \vec{k}_{p}-\vec{k}_{s}\right) $. After 
$m$ photons scattered into the $z$ direction and $n$ photons scattered in
the opposite direction, the atom occupies a state $\psi _{nm}$ with momentum 
$\hbar \left( m\vec{q}_{1}+n \vec{q}_{2}\right) $. During scattering,
the kinetic energy of the atom changes and the frequency of the scattered
photon is recoil shifted relative to the frequency $\omega $ of the pump
light by an amount specific to the transition $\psi _{n,m}$ $\rightarrow $ $%
\psi _{n^{\prime },m^{\prime }}$. With the cavity set to a fixed resonance
frequency one may thus selectively drive a particular transition by tuning $%
\omega $ to a value such that the scattered light is in resonance with the
cavity. Furthermore, several transitions can be driven simultaneously with a
pump laser that contains more than one frequency component. 

For studying matter wave subradiance we use the three atomic
momentum states with $m=0$ and $n=0,1,2$ (Fig. 1). The electric field of the
pump beam contains two frequency components $\omega $ and $\omega +\Delta $
with amplitudes $E_{0}$ and $E_{1}$ that resonantly drive the transition $A$
from $n=0$ to $n=1$ and transition $B$ from $n=1$ to $n=2$. The atoms are
initially prepared in the zero momentum state ($n=0$). With only one
frequency in the pump beam ($E_{0}>0$, $E_{1}=0$) all atoms quickly decay
into the intermediate state ($n=1$). Further decay into the state $n=2$ is suppressed by the cavity. If the second frequency
component is added ($E_{1}>0$), one would naively expect a complete two step decay into the ground state ($n=2$). In contrast, one observes a substantial fraction of atoms remaining in the states with $n=0$ and $n=1$. A complete decay into $n=2$ is efficiently suppressed and the
system evolves into a stationary subradiant state. Once the subradiant state has
settled, scattering is absent and the resonator contains no light despite
resonant pumping of the system on both transitions. 

We carry out the experiment with $N=250\,000$ atoms of a magnetically trapped $%
^{87}$Rb Bose-Einstein condensate in state $F=2,m_{F}=2$. The pump light is
generated by an amplified external cavity diode laser which is locked to a
transverse mode of the cavity (TEM$_{10}$) with an uncertainty of better
than $\SI{200}{\hertz}$. This reference mode has very little spatial overlap with the
atoms and thus does not interact with the atoms \cite{BuxAPB}. The frequency
of the pump light is red detuned relative to the frequency $\omega _{0}$ of
the $D1$ transition ($5s$, $F=2$, $m_{F}=2$ to $5p_{1/2}$, $F=2$, $m_{F}=2$)
by $\Delta _{a}=\omega -\omega _{0}=-\SI{100}{\giga\hertz}$. To create the two frequency
components of the pump light the output of the pump laser is split into two
beams. The frequency of each beam is shifted with an acousto-optic modulator
and the two beams are subsequently recombined with a beam splitter. By
adjusting the relative power of the two beams the amplitude ratio $%
\varepsilon =E_{1}/E_{0}$ of the two resulting optical frequency components
can be varied between $\varepsilon =0$ and $\varepsilon =3$. The pump light with a beam waist of $\omega_0=\SI{200}{\micro\meter}$ illuminates the atoms with a total intensity of about $\SI{1}{\watt\per\square\centi\meter}$. With an incident
angle of $\phi =148{{}^\circ}%
$ the recoil frequency amounts to $\omega _{r}=\left[ 2\hbar k_{p}\sin
\left( \phi /2\right) \right] ^{2}/\left( 2\hbar M\right) =2\pi \times\SI{13.6}{\kilo\hertz}$ with $M$ being
the mass of the atom. The pump light and the light in the resonator are
linearly polarized and oriented perpendicular to the plane of the resonator.
For this polarization the field decay rate of the light in the resonator of $\kappa =2\pi \times \SI{5}{\kilo\hertz}$ lies well below the recoil frequency $\omega
_{r}$. The round trip length of the resonator is $\SI{39}{\centi\meter}$ and the beam
waist at the position of the atoms $w_{0}=\SI{170}{\micro\meter}$. 
The volume of the fundamental mode amounts to $V=\SI{19}{\milli\meter^3}$ and the Purcell
enhancement is calculated to be $\eta _{P}=6F/(\pi k_{p}^{2}w_{0}^{2})=0.1$
with $F=94\,000$ being the finesse of the cavity. The single photon light shift amounts to about $\SI{0.02}{\hertz}$ and the atom-cavity system is far in the weak coupling regime. After the condensate is
placed inside the resonator the pump beam illuminates the atoms for a
variable exposure time and the power in the resonator mode is monitored by
observing the light that leaks out at one of the cavity mirrors with a
sensitive avalanche photodiode (see Fig.1). At the end of the exposure
time, the magnetic trap is turned off and the atomic momentum distribution
is observed by imaging the atomic cloud after $\SI{25}{\milli\second}$ of ballistic expansion.
Special care is taken to separate the population $\left| c_{n}\right| ^{2}$
of the coherent momentum states from the background of thermal atoms, induced by an unwanted broadband spectrum of the optical amplifier (see Supplemental Material \cite{supplement_mat}). This generates a slowly growing thermal cloud around the condensate. The total number $N$ of atoms in the three momentum states decreases correspondingly. This is taken into account in the numerical simulations below.

To trigger the superradiant decay into the subradiant state we inject resonant laser light into the resonator mode along the negative z direction. In a pure homogeneous condensate the decay may start from quantum fluctuations of the electromagnetic vacuum inside the resonator. In a real experiment, finite-size effects, residual atomic density fluctuations, light scattering from a thermal halo, and seismic noise may also play a role. Experimentally, it has turned out that a short seed pulse at the beginning of the experiment starts the decay in a controlled way. During the
seed pulse the circulating power inside the cavity is about 5 times larger
than the power of the superradiant pulse, which is observed when the system
relaxes into the subradiant state.

We numerically simulate the experimental observations with the equations of motion
in mean field approximation Ref. \cite{Piovella}%
\begin{eqnarray*}
\frac{dc_{n}}{dt} &=&i\omega _{n}c_{n}+g(\alpha a^{\ast }c_{n-1}-\alpha
^{\ast }ac_{n+1}) \\
\frac{da}{dt} &=&gN\alpha \sum_{n}c_{n}c_{n+1}^{\ast }-\kappa a+\eta
\end{eqnarray*}%
where $\omega _{n}=n\left( n\omega _{r}-\delta \right) $. The
pump cavity detuning $\delta =\omega -\omega _{c}$ contains the frequency $%
\omega $ of the pump component which drives transition $A$ and the resonance
frequency of the resonator $\omega _{c}$, which includes a time independent
correction due to the index of refraction of the atoms. The macroscopic
atomic wave function is expanded in momentum eigenstates $\psi \left(
z,t\right) =\sum c_{n}\left( t\right) \exp \left[ in\left( \vec{q_{2}}\vec{r}-\delta
t\right) \right] $. The expansion
coefficients $c_{n}$ are normalized according to $\sum \left| c_{n}\right|
^{2}=1$. The amplitude $a$ of the light mode is normalized such that $|a|^{2}$ equals the number of photons inside the resonator. The time dependent expression $\alpha \left( t\right)
=1+\varepsilon e^{-i\Delta t}$ contains the amplitude ratio $\varepsilon$ between the two
frequency components of the pump beam. The
frequency of the second component with amplitude $E_{1}$ is detuned by $%
\Delta $ relative to the first component. Finally, the coupling parameter $%
g=d^{2}\ E_{0}\sqrt{\omega /(8\hbar ^{3}\varepsilon _{0}V)}/\Delta _{a}$
contains the dipole matrix element $d$ of the rubidium $D1$ transition, the
amplitude of the first pump field component $E_{0}$ and the permittivity of free
space $\varepsilon _{0}$. The sum in the equation of motion is the structure factor that describes the coupling of light to the atomic density waves. Neglecting all terms with $n<0$ and $n>2$, the sum reduces to the two terms $c_{0}c_{1}^{*}$ and $c_{1}c_{2}^{*}$. These terms are proportional to the contrast of the atomic density waves. The temporal behavior of their complex phases directly translates in the velocity of the density waves. The last term in the equation for the light describes the resonantly injected seed light. The parameter $\eta$ denotes the amplitude of the injected light. The above-mentioned reduction of $N$ due to heating is taken into account by assuming an exponential decay for $N$ in the equations of motion with a decay constant of $\SI{1}{\milli\second}$. For simulating the observations we include all momentum states between $n=-5$ to $n=5$. The states with negative $n$ have
almost no effect. The population of the state with $n=2$ is slightly
affected by the states with $n>2$. In the experiment and the simulations
both transitions are resonantly driven, i.e., $\delta =\omega _{r}$ and $%
\Delta =2\omega _{r}$.

   \begin{figure}[htb]
        \begin{center}
            \includegraphics[width=3.3in]{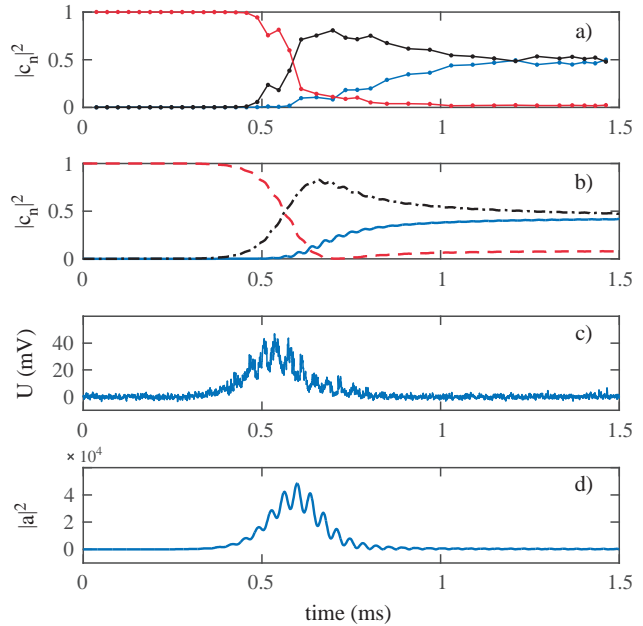}  
        \end{center}

        \caption {Relaxation into the subradiant state without seed pulse. (a) Observed
        relative population $\left| c_{n}\right| ^{2}$ of the momentum states with $%
        n=0,1,2$ (red, black, blue dots) for various holding times after turning on
        the pump beams. The amplitude ratio $\varepsilon =0.6$ and the total pump
        power amounts to $\SI{1.9}{\milli\watt}$. Each data point shows an average over six
        experimental cycles. (b) Simulation for the populations for the parameters of (a). (c) Power
        in the cavity mode during a typical cycle ($\varepsilon =0.6$)
        and (d) simulation.}
        
    \end{figure}

Figure 2(a) shows a typical decay into the subradiant state without applying a seed pulse for an amplitude
ratio $\varepsilon =0.6$. It takes about $\SI{500}{\micro\second}$ until the 
collapse into the subradiant state starts from noise. After about 
$\SI{1}{\milli\second}$ the system reaches a steady state. The dynamics is well
described by the simulation [FIG. 2(b)]. The decay is accompanied by a
superradiant light pulse as shown in Fig. 2(c) together with the simulation in
Fig. 2(d). The incomplete decay into the final state ($n=2$) is clearly observed. For this 
value of  $\varepsilon$ almost half the population is trapped in the intermediate state ($n=1$)
with only a small fraction of the atoms remaining in the zero momentum state.
Much larger fractions are observed for larger values of $\varepsilon$ (see Fig. 4).

   \begin{figure}[htb]
        \begin{center}
            \includegraphics[width=2.87in]{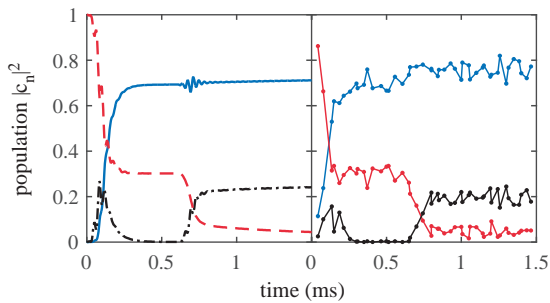}
        \end{center}

        \caption {Triggering the decay into the subradiant state with
        a seed pulse. Population  $\left| c_{n}\right| ^{2}$ of the
        momentum states with $n=0,1,2$ (red dashed, black dash-dotted, and blue solid
        line) for various holding times after turning on the pump beams. At $t=0$
        and $\SI{0.6}{\milli\second}$ seed light is injected into the cavity mode for $\SI{100}{\micro\second}$. The pump ratio $\varepsilon =1.8$ was set to zero immediately before
        the second seed pulse is applied. Simulation and observations are shown in
        the left and right subplot. Each experimental data point represents an
        average over six experimental cycles.}

    \end{figure}

Figure 3 shows the effect of a seed pulse for the decay into a subradiant state with $\varepsilon=1.8$. The delay of the initial superradiant pulse is almost completely eliminated. A further decay out of the subradiant state is shown by setting $\varepsilon=0$ and applying a second seed pulse after $\SI{600}{\micro\second}$. The remaining population in the initial state ($n=0$) now decays into the intermediate state ($n=1$) with the population of the final state ($n=2$) being unaffected.\\    

   \begin{figure}[htb]
        \begin{center}
            \includegraphics[width=2.8in]{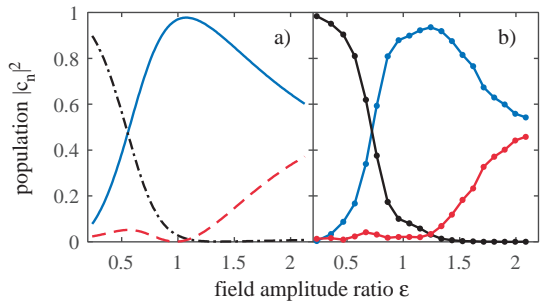}
        \end{center}

        \caption {Variation of the field amplitude ratio $\varepsilon $. Simulated relative population  $%
        \left| c_{n}\right| ^{2}$ of the momentum states with $n=0,1,2$ (red dashed, black dash-dotted, and blue solid line) after a holding time of $\SI{1.5}{\milli\second}$ for various
        amplitude ratio $\varepsilon $ [plot (a)] and measurement in plot (b). Each data point represents an average over 60 experimental cycles. The statistical error amounts to about 12\%. A series of subradiant states is observed with different steady populations in the initial state and the intermediate state. }

    \end{figure}

With initial seeding, we recorded the populations in the subradiant 
state for various amplitude ratios $\varepsilon $. The
observed behavior is well reproduced by the simulation. For small $\varepsilon$ all three momentum states are significantly occupied and the resulting atomic density consists of two waves that propagate at different speed. Because of the Doppler effect, each matter wave diffracts one of the two frequency components of the pump beam resonantly into the cavity. The two diffracted optical fields interfere destructively such that no light is scattered. This interpretation does not hold for larger $\varepsilon$ where the intermediate state is barely occupied. In this regime, an intuitive explanation can make use of rate equations that can be derived from the equations of motion for times longer than the cavity life-time \cite{Bux}. Different than the usual rate equations, here the transition rates are bosonically enhanced and are proportional not only to the population of the initial state but also to the population of the final state. With no occupation in the intermediate state the decay from the initial zero momentum state is thus effectively suppressed. Driving transition $B$ depopulates the intermediate state and consequently inhibits transition $A$. This is only possible after some population has accumulated in the final state during the formation of the subradiant state. Otherwise also transition $B$ would be suppressed. As expected for three level systems, decaying into the subradiant state requires different rates for transition $A$ and $B$ (see Supplemental Material \cite{supplement_mat}). In fact for $\varepsilon=1$ almost all atoms decay into the final state. Since in the simulation also states up to $n=5$ are taken into account, small deviations from the pure three level approximation are to be expected. 

Note that in the experiment we only observe the classical properties of the subradiant states as captured by the mean field approximation. However, as the toy model in the supplemental material already shows, subradiant states are entangled states that cannot be fully described as product states of three level systems. In fact, in the here investigated regime for $\varepsilon <1+\sqrt{2}$, a quantum analysis on the particle level predicts entanglement between the three momentum states and non classical second order correlation functions \cite{Piovella}. The direct observation of these particle correlations requires more sensitive detection methods to be developed. Our experiment mainly demonstrates the accessibility of this regime.

In summary, we have observed the formation of stationary subradiant states in
an ensemble of three level systems formed by atomic momentum states. Recently, suppression of superradiance has also been observed with a Bose-Einstein condensate pumped by two counterpropagating laser beams  \cite{Dimitrova}. Although this suppression is not based on destructive interference of emitted radiation a more detailed comparison with atoms in cavities might still be helpful to sharpen the understanding of subradiance in the context of atomic quantum gases. For the
future it will be interesting to study quantum correlations of the
subradiant state \cite{Piovella} and explore possible applications for
storage of quantum information. Techniques for efficient momentum selective single atom detection
would be desirable. A possible approach could be state selective ionization as 
investigated in Ref. \cite{Gunther}. 

We acknowledge helpful discussion with N. Piovella and A. Hemmerich. The work is funded by the Deutsche Forschungsgemeinschaft.\\

\end{document}